# Three-dimensional bulk electronic structure of the Kondo lattice CeIn$_3$ revealed by photoemission

Yun Zhang[1,2*], Haiyan Lu[1], Xiegang Zhu[1], Shiyong Tan[1], Qin Liu[1], Qiuyun Chen[1], Wei Feng[1], Donghua Xie[1], Lizhu Luo[1], Yu Liu[4,5], Haifeng Song[4,5], Zhengjun Zhang[3], Xinchun Lai[1*]

[1]*Science and Technology on Surface Physics and Chemistry Laboratory, Mianyang 621907, China*

[2]*Department of Engineering Physics, Tsinghua University, Beijing 100084, China*

[3]*School of Materials Science and Engineering, Advanced Materials Laboratory, Tsinghua University, Beijing 100084, China*

[4]*Laboratory of Computational Physics, Institute of Applied Physics and Computational Mathematics, Beijing 100088, China*

[5]*CAEP Software Center for High Performance Numerical Simulation, Beijing 100088, China*

We show the three-dimensional electronic structure of the Kondo lattice CeIn$_3$ using soft x-ray angle resolved photoemission spectroscopy in the paramagnetic state. For the first time, we have directly observed the three-dimensional topology of the Fermi surface of CeIn$_3$ by photoemission. The Fermi surface has a complicated hole pocket centred at the Γ-Z line and an elliptical electron pocket centred at the R point of the Brillouin zone. Polarization and photon-energy dependent photoemission results both indicate the nearly localized nature of the 4$f$ electrons in CeIn$_3$, consistent with the theoretical prediction by means of the combination of density functional theory and single-site dynamical mean-field theory. Those results illustrate that the $f$ electrons of CeIn$_3$, which is the parent material of CeMIn$_5$ compounds, are closer to the localized description than the layered CeMIn$_5$ compounds.

## Introduction

Heavy fermion (HF) compounds CeMIn$_5$ (M=Co, Rh, Ir) have attracted much attention in the last decade because of their novel properties.[1-6] For example, the 4$f$ electrons of CeCoIn$_5$ go through a transition from the localized state to the itinerant state and begin to participate in the modification of the Fermi surface (FS) at low temperatures,[1,7,8] resulting in an abnormal enhancement of the electron mass. Under certain conditions, those heavy $f$ electrons condense into Cooper pairs, e.g., CeCoIn$_5$ exhibits the highest superconductivity (SC) temperature recorded in Ce-based HF compounds.[2,3,9,10] However, there are still many unresolved questions regarding this system. First, there is not an explicit definition of the crossover line from the localized $f$ electrons to the itinerant $f$ electrons state. Second, the SC of CeCoIn$_5$ cannot be explained by the BCS theory.[1,2,11,12] As layered compound, the structure of CeMIn$_5$ is comprised of alternating layers of CeIn$_3$ and MIn$_2$. The three-dimensional (3D) component CeIn$_3$ in CeMIn$_5$ contributes all the $f$ electrons and can be viewed as adding an effective positive pressure on the CeIn$_3$ crystal at ambient pressure.[13,14] Further investigation of the electronic structure of CeIn$_3$ is of particular importance for understanding the nature of CeMIn$_5$. However, unlike the two-dimensional (2D) CeMIn$_5$ compounds,[7,8] the electronic structure of CeIn$_3$ is 3D. Consequently, the study of CeIn$_3$ also provides an opportunity to study the influence of the layered structure on the properties of $f$ electrons in this system.

To investigate the properties of $f$ electrons in CeIn$_3$, many experiments have been previously performed. Transport,[15] optical conductivity spectra,[16] and inelastic neutron scattering[17-19] results all indicate the existence of HF at low temperature, even inside the antiferromagnetic (AFM) phase below 10K. Quantum oscillation measurements[20,21] reveal that the HF states occupy only a small portion of the FS. Polycrystal photoemission measurements[22], angular correlation of the electron-positron annihilation radiation[23], and de Haas-van Alphen (dHvA)[24] measurements all reveal the localized nature of $f$ electrons at ambient pressure. Photoemission spectroscopy is a powerful tool to detect the reconstruction of the electronic structure and is often used to judge the nature of $f$ electrons.[25-28] Photoemission spectroscopy can even be used to detect the SC energy gap in HF systems[29]. However, angle-resolved photoemission spectroscopy (ARPES) studies of $CeIn_3$ have not been previously performed due to the difficulty of cleaving the sample and the complicated 3D electronic structure of $CeIn_3$.

In this work, the electronic structure of $CeIn_3$ in the paramagnetic state is characterized using the soft x-ray ARPES technique for the first time. The 3D topology of the FS of $CeIn_3$ is observed. The FS has a complicated hole pocket centred at the Γ-Z line and an elliptical electron pocket centred at the R point of the Brillouin zone (BZ). Photoemission results indicate the nearly localized nature of the 4$f$ electrons in $CeIn_3$, consistent with the theoretical predictions of the localized assumption. These results can help us to understand the behaviours of $f$ electrons in $CeIn_3$ and the derived $CeMIn_5$ systems.

# Results

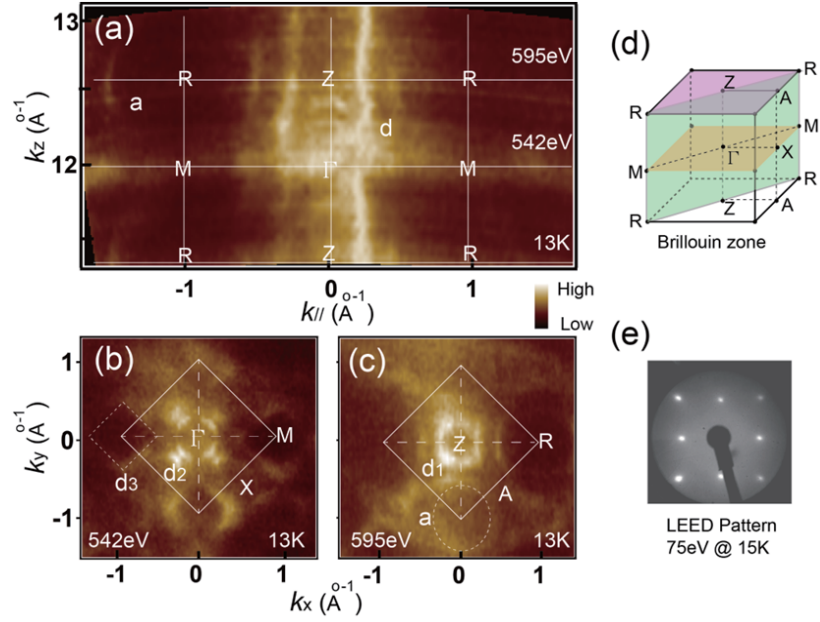

FIG. 1. 3D FS map of CeIn$_3$. (a) FS map in the $k_z$-$k_{xy}$ plane at the Fermi energy integrated over a window of [E$_F$-100 meV, E$_F$+100 meV]. Photon energies are varied from 480 eV to 650 eV at a step of 4 eV. 2D FS maps in the $k_x$-$k_y$ plane at the Fermi energy: (b) integrated over a window of [E$_F$-80 meV, E$_F$+80 meV] for a photon energy of 542 eV ($k_z$~0) and (c) integrated over a window of [E$_F$-130 meV, E$_F$+130 meV] for a photon energy of 595 eV ($k_z$~π). (d) Bulk BZ of CeIn$_3$ and the high symmetry points. The green, yellow, and purple planes correspond to the FS slices in Figs. 1(a), (b), and (c), respectively. (e) LEED pattern of the obtained CeIn$_3$ (001) surface. The bright spots in the square lattice reflect the pristine 1×1 surface.

**Fermi surface mapping:**

The topology of the FS of CeIn$_3$ is presented in Fig. 1. Figure 1 (a) shows the slice of the FS in the $k_z$-$k_{xy}$ plane, observed by hν-dependent ARPES. Although the FS contours contain complicated features, we can obtain the symmetry of the electronic structure along $k_z$ direction. The slices of the FS observed by 542eV and 595eV photon energies correspond to the centre and boundary of the BZ, respectively. Besides, a complicated band structure $d$ centred at the Γ-Z line and an elliptical shaped pocket $a$ centred at R point are observed. To obtain a better understanding of the FS, two selected photon energies are adopted to characterize the FS contours in $k_x$-$k_y$ plane. Figures 1 (b) and 1 (c) are the $k_x$-$k_y$ maps at $k_z\sim0$ (h$\nu$=542 eV) and $k_z\sim\pi$ (h$\nu$=595 eV), respectively. At the centre of the BZ in Fig. 1(b), the square-like pocket $d_3$ at the M point and the double-ring band $d_2$ centred at the Γ point are observed. The spectral weight of $d_2$ is quite strong in the Γ-X direction. At the boundary of the BZ in Fig. 1(c), a square structure $d_1$ around the Z point and an elliptical structure $a$ centred at the R point are displayed.

The topology of the FS is highly 3D in nature and agrees with the dHvA experiment and the full-potential linear augmented plane wave calculations results,[24] which exhibit a complicated structure (denoted as $d$) centred at the Γ-Z line and an ellipsoid-like structure around the R point. Figure 1 (e) shows the low energy electron diffraction (LEED) pattern of the obtained surface. A clean 1×1 pattern of the (001) surface is observed.

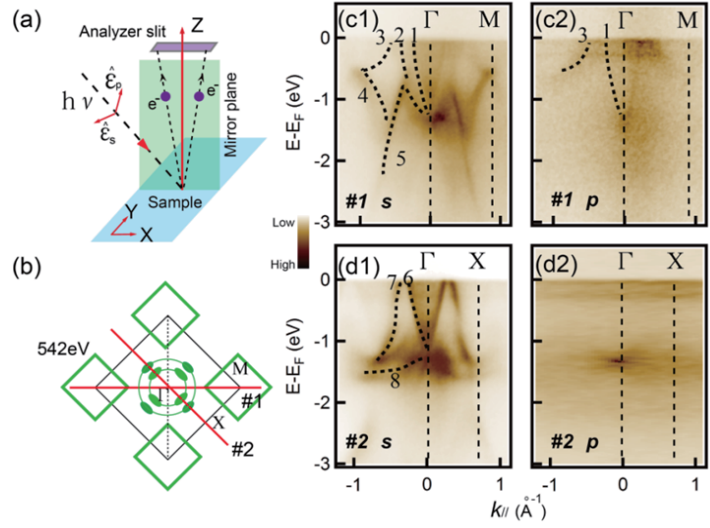

FIG. 2. Polarization dependent valence band structure of CeIn$_3$. (a) Experimental schematic of polarization dependent ARPES. (b) The centre of the BZ of CeIn$_3$ and the locations of the momentum cuts. (c1) and (d1) Photoemission intensity plots obtained using 542 eV *s*-polarized light along the Γ-M and Γ-X directions, respectively. (c2) and (d2) Photoemission intensity plots obtained using 542 eV *p*-polarized light along the Γ-M and Γ-X directions, respectively. The black dashed lines represent the highest intensity of the ARPES results.

**Valence band structure:**

An important goal in the study of a HF system is to understand the properties of *f* electrons. Previously, the polarized ARPES technique has been conducted to resolve the multi-orbit nature of the 3*d* states in iron-based superconductors and has proven to be an effective method to distinguish the different orbits of 3*d* electrons.[30] The HF compound YbRh$_2$Si$_2$ has also been studied using different light polarizations.[31,32] Some states in YbRh$_2$Si$_2$ are proven to possess dichroic effects, indicating that electronic states with the same symmetry can be identified by their similar response to a change of light polarization. Therefore, to investigate the possible multi-orbit properties of the *f* electrons in the Ce-based HF system, we have performed polarization-dependent ARPES measurements in CeIn$_3$.

The experimental valence band structures of CeIn$_3$ are displayed in Fig. 2. The electron-like band *1*, band *2* centred at the Γ point and band *3* around the M point can be clearly observed in Fig. 2(c1), forming the double-ring structure $d_2$ and the hole pocket $d_3$ observed in Figs. 1(b). Band *4* is located at the bottom of band *3*. Band *5* extends to high energy and nearly connects with band *1* around the Γ point at approximately 1 eV binding energy (BE). Three bands (bands *6*, *7* and *8*) are observed along the Γ-X direction in Fig. 2(d1). The electron-like band *6* and hole-like band *7* are adjacent at the Fermi level. Band *8* is located at approximately 1.2eV BE. Except for the band structures discussed above, two nearly non-dispersive bands located at E$_F$ and 300meV BE are observed in Figs. 2(c1) and (d1). They have much weaker intensity weight than the other conduction bands and can be observed more clearly in Fig. S1 in the SI. The origin of the non-dispersive bands could be Kondo effect or spin-polaron effect,[33,34] which both lead to the many-body resonance in the electron density of states near E$_F$. However, the temperature dependent resistivity of CeIn$_3$ can be described by a function ρ~-lnT beyond the coherent temperature,[15,35] indicating Kondo physics takes effect. Therefore, the two non-dispersive bands correspond to the $4f^1_{5/2}$ state and its spin orbit coupling (SOC) sideband $4f^1_{7/2}$.[8,22,25] The two flat bands are observed more clearly, as shown in Fig. 2(c2) and (d2), where significant changes occur when light is changed to *p*-polarized light. Bands *2*, *4,* and *5* along Γ-M and bands *6* and *7* along Γ-X disappear. Bands *1* and *3* and the two non-dispersive *f* bands remain. The suppression of the intensity of the conduction states is presumably due to the effect of the relative spatial orientations of the electronic states with the polarization vector of the light. The phenomena, by using *p*-polarized light, might be applied to other Ce-based HF compound, even with a low photoemission cross-section, as the suppression of non-*f* states makes *f* states in ARPES intensity plots clearer and purer.

Another unresolved problem in CeIn$_3$ is determining why the *f* bands can be clearly observed by off-resonance photoemission spectroscopy in Fig. 2. In contrast, for many other HF systems, *f* bands can only be observed using on-resonance photoemission.[8,25] We have studied the photoemission cross sections of different orbits of Ce and In as a function of photon energy. Ce (4*f*, 5*d*, 6*s*) and In (5*s*, 5*p*) states were considered as valence states.[36] In the photon energy range from 500 eV to 900 eV, the total photoemission cross section is mainly contributed by 4*f* electrons. The contribution from Ce (5*d*, 6*s*) and In (5*s*, 5*p*) are approximately one order-of-magnitude smaller.[37]

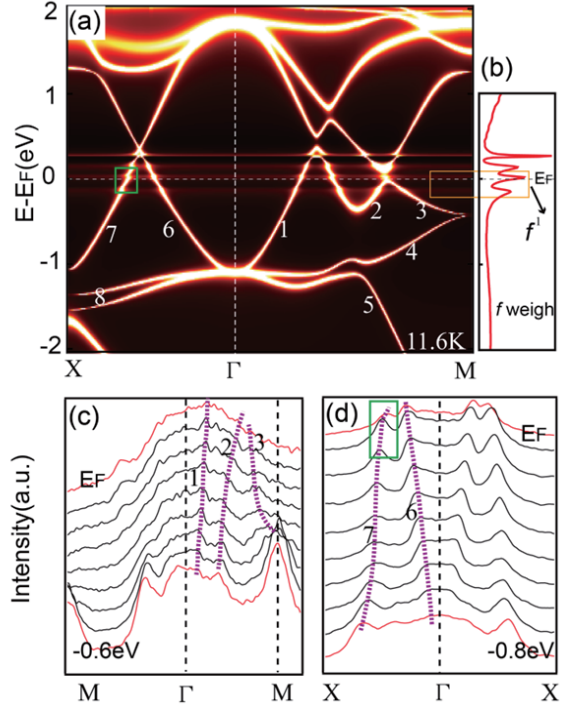

FIG. 3. (a) The band structure calculated using the DFT+DMFT approach for CeIn$_3$. (b) Partial density of states for the Ce-4$f$ states of CeIn$_3$ using DFT+DMFT. (c) and (d) MDCs display along Γ-M and Γ-X directions at $k_z$~0, respectively. The purple dotted lines are guides to eye, tracing the peaks of the MDCs. The orange rectangle marks the $f$ states. The green rectangles mark the renormalization of band 7.

**Calculations:**

The comparison between the experimental band structures with the calculation results is an effective means to judge the properties of the $f$ electrons. On the other hand, density functional theory merged with the single-site dynamical mean-field theory (DFT+DMFT) is probably the most powerful established method to study the electronic structures of strongly correlated materials. DFT+DMFT has been successfully applied in the studies of many HF systems, such as the temperature-dependent localized-itinerant transition in CeIrIn$_5$.[4,5] We compare the ARPES results with the DFT+DMFT band calculations of $CeIn_3$ in Fig. 3. First, all the valence bands *1-8* in Fig. 2 and Fig. 3(c-d) can be clearly duplicated in the calculation results, as are the very flat and non-dispersive *f* bands located at $E_F$ and the vicinity. The flat *f* bands appear to be constant and non-dispersive over all angles and have much weaker intensity weight than those of the other conduction bands, consistent with our experimental results in Fig. 2 and Fig. S1. This phenomenon does not agree well with the case of $CeIrIn_5$,[5] the *f* states of which have strong intensity weight and form the coherent peak on $E_F$. Besides, the *f* states of $CeIrIn_5$ show obvious dispersions around $E_F$, induced by the strong hybridization between *f* states and conduction states. This indicates that the *f* electrons of $CeIn_3$ have different behaviours compared with those of $CeIrIn_5$, which have itinerant *f* states at low temperature. Second, a small renormalization of conduction band *7* on $E_F$ is displayed in both the calculation results in Fig. 3(a) and the MDCs in Fig. 3(d). This should be induced by the interaction between *f* states and the band *7*. However, the interaction is so weak that the coherent peak does not form. In fact, if an additional pressure is added on $CeIn_3$, the interaction will be greatly enhanced. Band *7* will have obvious band bending and form the coherent peak on $E_F$ together with the hybridized *f* band.[38] Through comparison of the experimental bands with the calculated results, we can conclude that the experimental band structure of $CeIn_3$ at ambient pressure and low temperature can be described by the nearly localized model.

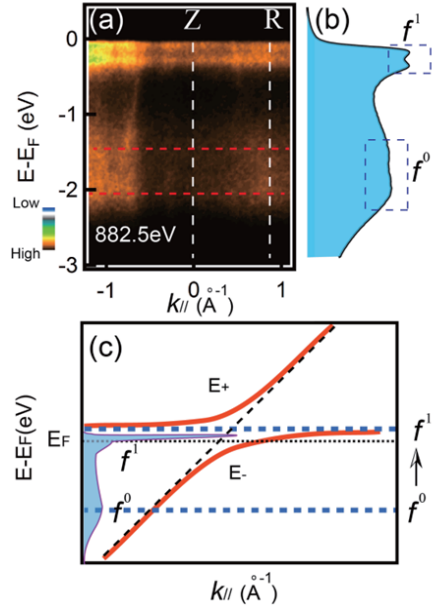

FIG. 4. On-resonance photoemission intensity plot of CeIn$_3$. (a) Photoemission intensity plot along the Z-R direction. The red dashed lines mark the two flat bands at -1.4 eV and -2 eV. (b) Angle-integrated photoemission spectroscopy of the ARPES result in Fig.4 (a). (c) Schematic of the Anderson lattice model. The dashed lines are the dispersive valence band and the non-dispersive $f$ bands. The dotted line indicates the location of $E_F$. For a finite hybridization, an energy gap opens, and two $k$-dependent branches E+ and E- form. The lower hybridized band E- crosses $E_F$ and forms the heavy particles. The blue shaded area is the angle-integrated photoemission spectroscopy curve within the hybridized model.

**On-resonance photoemission:**

To clarify the explicit band structures of the $f$ electrons, figure 4 (a) shows on-resonance photoemission intensity plot of CeIn$_3$. The dense and non-dispersive $f$ bands emerge near $E_F$ and ~300meV BE, enhanced by the 3$d$-4$f$ excitation. It is well known that the $f$ band at $E_F$ is the tail of the Kondo resonance (KR) peak, corresponding to the Ce-4$f^1_{5/2}$ final state. The peak at 300meV BE is the SOC sideband and corresponds to the Ce-4 $f^1_{7/2}$ final state. Although the Ce-4 $f^1_{5/2}$ band of CeIn$_3$ is intersected by the conduction band at E$_F$, it does not show any dispersion and does not open an energy gap to form the *k*-dependent bands E-, as shown in Fig.4 (c). In the Periodic Anderson Model (PAM),[39] if the *f* electrons are itinerant and have periodicity in the lattice, it should form dispersive KR peaks crossing E$_F$ in unoccupied states, as shown in the pressure-induced itinerant state for CeIn$_3$.[38] However, no crossing of *f* electron bands and no energy gap for CeIn$_3$ are observed at ambient pressure in Fig. 4 (a).

From another aspect, as shown in Fig. 4(b), except for the $f^1$ final states near E$_F$, the non-dispersive structures at approximately 1.4eV and 2eV BE are also visible. These arise from pure charge excitations of the trivalent Ce ion ($4f^1 \rightarrow 4f^0$) and are referred to the ionization peaks.[22] The two *f* bands have been observed in the Ce-termination surface in CeRh$_2$Si$_2$,[40] which exhibits the nature of localized *f* electrons. Usually, if the hybridization strength is strong, then the $f^0$ peak is much weaker than the $f^1$ final state, as shown in the blue shaded area in Fig.4 (c). If the hybridization strength is weak, then the intensity of $f^0$ peak is strong,[22,41] as demonstrated by recent calculation results using the DFT+DMFT approach by H.Y.L.[38] H.Y.L *et al.* indicate that the 4*f* electrons of CeIn$_3$ will undergo a localized-itinerant transition under pressure, consistent with the dHvA results.[24] At ambient pressure, the 4*f* electrons of CeIn$_3$ are localized, and the intensity of $f^1$ final state is comparable with that of $f^0$ peak. With the crystal volume decreasing under pressure, the intensity of the $f^1$ final state quickly increases, and the intensity of $f^0$ peak decreases. When the crystal volume of CeIn$_3$ decreases to 65% of the volume at ambient pressure, the $f^1$ peak is quite strong, and the $f^0$ peak nearly disappears. As the intensity of the $f^0$ peak is comparable with that of $f^1$ final state in CeIn$_3$ in Fig. 4 (b), the angle-integrated photoemission spectroscopy result is also consistent with the calculation result of CeIn$_3$ under ambient pressure. In summary, the *f* electrons of CeIn$_3$ are close to the description of the nearly localized model, similar to the angle-integrated photoemission spectroscopy results of *f* localized compounds CeRhIn$_5$ and CeRh$_2$Si$_2$.[40,42]

## Discussion

We have compared the ground states between CeIn$_3$ and CeMIn$_5$ to illustrate the different properties of the *f* electron. In fact, there are three possible ground states in the HF systems based on the Doniach phase diagram[43] at low temperature: i) the mixed valence ground state with extremely large hybridization strength between the *f* electrons and the conduction electrons and 0<n$_f$<1, where n$_f$ represents the occupation number of *f* electrons; ii) the magnetic ground state with weak hybridization strength and n$_f$≈1; iii) Non-Fermi liquid (sometimes SC) ground state with strong spin fluctuations, located between the two states discussed above.

CeCoIn$_5$ and CeIrIn$_5$ become superconducting at low temperature, implying that they are located in the SC region in the Doniach phase diagram and that they have relatively strong hybridization strength. In contrast, CeIn$_3$ has the antiferromagnetic ground state with much weaker hybridization strength. Although CeRhIn$_5$ is also an antiferromagnetic compound below 3.8K, the CeIn$_3$ units in it can be viewed as adding a pressure of approximate 1.4GPa to the CeIn$_3$ compound at atmospheric pressure.[14] For CeIn$_3$, the hybridization strength is enhanced under pressure.[16,24] This indicates that CeRhIn$_5$ has a larger hybridization strength than that of CeIn$_3$. These results agree well with our ARPES results of CeIn$_3$, regarding that the 4$f$ electrons of CeIn$_3$ are a nearly localized type.

Now the DFT+DMFT calculation results, our ARPES results, angular correlation of the electron-positron annihilation radiation[23] and dHvA[24] results all support the view that the 4$f$ electrons of CeIn$_3$ at ambient pressure are nearly localized. However, the optical conductivity results,[16] transport[15] and inelastic neutron scattering[17-19] measurements hold the opposite view. Why are the conclusions of different references on CeIn$_3$ totally different? First, from our DFT+DMFT calculations and experimental results in Fig. 3, the interaction between band 7 and $f$ band really exists, although the intensity of the interaction strength is too small to form the coherent peak, implying that most of the $f$ electrons are localized and a small portion of the $f$ electrons tend to be itinerant, but incompletely. Second, references 20 and 21 prove that the hybridized $f$ holes exist at low temperature in CeIn$_3$. However, the $f$ holes just occupy a small portion of the FS and are not along the high symmetry directions in BZ. Such a small proportion may make some techniques hard to detect them. However, this situation will change under pressure. More $f$ electrons begin to participate in the modification of the FS and the collective behaviors of $f$ electrons make the system itinerant and heavy under pressure.[33] Based on the above discussions, we propose that most of the $f$ electrons of CeIn$_3$ stay localized at ambient pressure and the situation can be changed by means of adding additional pressure.

In summary, the electronic structure of CeIn$_3$ in the paramagnetic state was characterized using the soft x-ray ARPES technique. 3D FS of CeIn$_3$ was revealed. The FS has a complicated hole pocket *d* centred at the Γ-Z line and an elliptical electron pocket *a* centred at the R point of the BZ. The photoemission results and the calculated results all indicate a nearly localized nature of the 4*f* electrons in CeIn$_3$ at ambient pressure.

## Methods

High-quality single crystals of CeIn$_3$ were grown using the self-flux method.[44] The fresh and smooth surfaces were obtained by performing cycles of Ar$^+$-ion sputtering and annealing with a base pressure better than 3×10$^{-10}$ mbar after the surfaces are polished in the atmosphere. The polarization and photon-energy dependent soft x-ray ARPES experiments were performed at the ADDRESS station of the Swiss Light Source facility. The soft x-ray ARPES spectra were obtained using a PHOIBOS-150 photoelectron analyser.[45] The combined energy resolution is 90 meV or better, and the angle resolution is 0.1°. The base pressure of the ultra-high vacuum system was below 5×10$^{-11}$ mbar during the entire measurement. The samples were kept at T=13 K in the ARPES measurements. Unless a particular explanation is given, all the data are taken using *s*-polarized light.

The calculation method is the density functional theory merged with the single-site dynamical mean-field theory (DFT+DMFT) that combines the first-principles aspect of DFT with the non-perturbative many-body treatment of local interaction effects in DMFT. The method used in this paper is introduced in detail in the literature.[38] All of the calculations were conducted at the inverse temperature β=1000 (T=11.6 K), which is comparable with the experimental temperature. Here we adopted U=6.2 eV and J=0.7 eV, where U is the Coulomb interaction strength and J the Hund's exchange parameter.

*Corresponding author: E-mail:

yun-zhang13@mails.tsinghua.edu.cn (Y.Z.)

laixinchun@caep.cn (X.C.L.)

**Acknowledgements**

We thank Dr. Federico Bisti and Vladimir Strocov for providing experimental support at the Swiss Light Source(SLS). We gratefully acknowledge the helpful technique support from Prof. D.L.Feng and H.C.Xu. This work was supported by the Foundations for Development of Science and Technology of China Academy of Engineering Physics (No.2012A0301014), the National Natural Science Foundation of China (No.11304291, No. 11176002 and No. 11504061) and the National High Technology Research and Development Program of China under Grant 2015AA01A304.

**Author contributions**

Y.Z., X.G.Z., S.Y.T. and Q.Y.C. conducted the ARPES experiments. H.Y.L. conducted the calculations. Y.Z. and D.H.X. grew the $CeIn_3$ single crystals. W.F. and L.Z.L. provided the sample analysis. Z.J.Z. and X.C.L. designed the project. Y.Z., Q.L., Y.L., H.F.S., Z.J.Z. and X.C.L. prepared the manuscript. All authors have read and approved the final version of the manuscript.

**Competing financial interests**

The authors declare no competing financial interests.